\documentclass[fleqn,10pt]{wlscirep}
\usepackage[utf8]{inputenc}
\usepackage[T1]{fontenc}
\usepackage{graphicx}
\usepackage{color}
\usepackage{hyperref}
\usepackage{amssymb,amsthm}
\usepackage{array,bm}
\usepackage{comment,enumitem,setspace,ulem}
\usepackage{algorithm,algpseudocode}
\usepackage[misc]{ifsym}
\usepackage{xurl}

\renewcommand{\v}[1]{{\bm #1}}
\renewcommand{\cal}[1]{\mathcal #1}
\renewcommand{\b}[1]{\mathbf #1}

\newtheorem{proposition}{Proposition}
\newtheorem{lemma}{Lemma}

\algrenewcommand\algorithmicrequire{\textbf{Input:}}
\algrenewcommand\algorithmicensure{\textbf{Output:}}
\newcommand{\nn}{\nonumber \\}
\newcommand{\norm}[1]{\| #1 \|}
\newcommand{\inner}[2]{\left<#1,#2\right>}
\newcommand{\ket}[1]{\left| #1 \,\right\rangle}

\newcommand{\Z}{\mathbb{Z}}
\newcommand{\R}{\mathbb{R}}

\title{Assessing the feasibility of quantum learning algorithms for noisy linear problems}

\author[1$\dag$]{Minkyu Kim}
\author[1*]{Panjin Kim}
\affil[1]{The Affiliated Institute of ETRI, Daejeon 34044, Korea}

\affil[$\dag$]{mkkim@nsr.re.kr}
\affil[*]{pansics@gmail.com}

\keywords{Quantum algorithm, Quantum Fourier Transform, Learning with errors, Machine learning, Bernstein-Vazirani algorithm}

\begin{abstract}
Quantum algorithms for solving noisy linear problems are reexamined,
under the same assumptions taken from the existing literature.
The findings of this work include on the one hand extended applicability of
the quantum Fourier transform to the ring learning with errors problem which
has been left open by Grilo et al., who first devised a polynomial-time
quantum algorithm for solving noisy linear problems with quantum samples.
On the other hand, this paper also shows there exist efficient classical algorithms for short integer solution and size-reduced learning with errors problems if the quantum samples used by the previous studies are given.

\end{abstract}
\begin{document}

\flushbottom
\maketitle
%
%
\thispagestyle{empty}

\section*{\label{sec:introduction}Introduction}

In designing algorithms, setting underlying
assumptions on what one can and cannot do is a crucial
starting point.
Indeed in a study of quantum algorithms, seemingly
innocent assumptions sometimes become the center of
controversies, for example one in a blackbox query~\cite{BHT98,CNS17},
in a state preparation~\cite{Aar15}, or on quantum random
access memory~\cite{JR23}.
Considering that the field is not in the mere proof-of-concept
stage, it is a necessary task to examine such an assumption
and evaluate its plausibility.

One of the rapidly growing fields in quantum technology
is machine learning where researchers try to make the best
use of physical phenomena in studying the patterns in data.
Following the pioneering work on the quantum approach to
learning parity with noise (LPN) problem by Cross et al.~\cite{CSS15}, 
an important milestone was achieved in 2019 by Grilo et al.~\cite{GKZ19}
where the authors have proven that learning with errors (LWE)
problem is efficiently solvable by using Bernstein-Vazirani (BV)
algorithm~\cite{BV93}, if a certain form of a superposed data
sample is available.
The paper also investigated the applicability of the
learning algorithm to LPN,
ring learning with errors (RLWE), learning with rounding,
and short integer solution (SIS) problems, although not all
were successful.
The algorithm was further revised in 2022 by utilizing the
divide-and-conquer strategy, basically dealing with component-wise
problems~\cite{Song22a}.

As clearly noted by the authors of the paper~\cite{GKZ19}, it does not
directly mean the end of some lattice based cryptography, as
it is not clear 
how such a sample can be prepared.
At the time of writing, there has been no known
way to efficiently prepare such samples, but
on the other hand, there has been no mathematical proof that
an efficient way is nonexistent.
To sum up, it is too early to be optimistic (or pessimistic) on
the plausibility of the learning algorithms along the line.


Nevertheless, we have noticed in some occasions specific to two papers~\cite{GKZ19,Song22a},
certain assumptions allow polynomial-time classical algorithms 
questioning their soundness, or the negative prospect 
given by the authors turns out to be feasible.
This work is summarized as follows:

\begin{itemize}[leftmargin=*]
  \item
  The LWE algorithm with quantum samples is extended to solve RLWE problem,
  which has been deemed to be infeasible under the error model by
  Bshouty and Jackson~\cite{GKZ19,BJ95}.
  \item
  It is shown that the quantum sample assumption in the SIS algorithm
  is stronger than that in LWE 
  in the sense that a polynomial-time classical algorithm 
  can be devised given the same sample introduced by the previous work.
  
  \item
  Divide-and-conquer algorithm takes advantage of
  the component-wise approach to the secret vector,
  assuming size-reduced quantum samples~\cite{Song22a}.
  We develop classical algorithms for the LWE problem given
  the same samples.
\end{itemize}

In the following, background information is covered prioritizing introduction to the LWE algorithm with quantum samples.
New findings are then followed.


\section*{\label{sec:preliminaries}Preliminaries}

This section first introduces cryptographic hard problems related with noisy linear learning,
and reviews the (generalized) Bernstein-Vazirani (BV) algorithm~\cite{BV93}
for solving the LWE problem with quantum samples.
Readers are assumed to be familiar with Bra-ket notation
and the elementary quantum computation.

The following notation is used throughout the paper.
\begin{itemize}[leftmargin=*]
  \item
  For $q> 2$, $\Z_q := [-\frac{q}{2},\frac{q}{2}) \cap \Z$.
  For $q=2$, $\Z_2 := \{0,1\}$.
  We also define $\Z^+ = \{ a \in \Z : a > 0\}$.
  \item
  An inner product of two vectors $\v a = (a_0, a_1, \ldots , a_{n-1})$,
  $\v b = (b_0, b_1, \ldots , b_{n-1})$ is defined as follows:
  \begin{align*}
  \inner{\v a}{\v b} = a_0 b_0 + a_1 b_1 + \cdots + a_{n-1} b_{n-1} \enspace.
  \end{align*}
  \item
  Let $x \leftarrow \chi$ denote  a variable $x$ taking a value according to the probability distribution $\chi$,
  and let $x_0,\ldots,x_{l-1}\stackrel{i.i.d.}{\longleftarrow} \chi$ denote that
  $x_0,\ldots,x_{l-1}$ are sampled independently according to the probability distribution $\chi$.
  Let $\mathcal{U}(S)$ denote the uniform distribution on a set $S$.
  Let $\mathcal{B}_\eta$ denote the Bernoulli distribution with parameter $\eta \in [0,1/2)$
  so that $\Pr[x = b  : x \leftarrow \mathcal{B}_\eta] = b \!\cdot\! \eta + (1-b) \!\cdot\! (1-\eta)$
  for $b \in \{0,1\}$.
  \item
  Let
  $\mathsf{poly}(x_1,\ldots,x_m) = O(x_1^{c_1} \cdots x_m^{c_m})$ for some constants $c_1,\ldots,c_m$.
\end{itemize}

\subsubsection*{Hard problems}

Here we describe cryptographic hard problems
such as LPN, LWE, RLWE, and SIS.
The hardness of these problems are guaranteed in the following sense:
LPN problem can be seen as the average-case analogue
of the decoding random linear codes which is NP-complete~\cite{BMV78},
and LWE, RLWE, SIS problems are known to be
at least as hard as certain lattice problems in worst-case~\cite{Reg05,BLPRS13,LPR10,Ajt96,MR07}
when they are properly parameterized.

\vspace{2pt}
\noindent \textit{Learning parity with noise.}
Let $\v s \in \{0,1\}^n$ be a secret vector
and let $\eta \in [0,1/2)$ be a noise parameter.
An LPN sample is given by
\begin{align*}
(\v x, \inner{\v s}{\v x} + e \bmod 2) \in \{0,1\}^n \times \{0,1\} \enspace,
\end{align*}
where $\v x \leftarrow \mathcal{U}(\{0,1\}^n)$ and
$e \leftarrow \mathcal{B}_\eta$.
The (search) LPN problem asks to find the secret vector $\v s \in \{0,1\}^n$,
given such LPN samples.

\vspace{2pt}
\noindent \textit{Learning with errors.}
The LWE problem is a generalization of the LPN problem.
Let $\v s \in \Z_q^n$ be a secret vector where $q \in \Z^+$,
and let $\chi_{\textup{\tiny lwe}}$ be a probability distribution on $\Z$,
called error distribution.
An LWE sample is given by
\begin{align*}
(\v a, \inner{\v s}{\v a} + e \bmod q)
\in \Z_q^n \times \Z_q \enspace,
\end{align*}
where $\v a \gets \mathcal{U}(\Z_q^n)$
and $e \gets \chi_{\textup{\tiny lwe}}$.
The (search) LWE problem asks to find $\v s$,
given such LWE samples.

\vspace{2pt}
\noindent \textit{Ring learning with errors.}
The RLWE problem is in general defined on a ring of
integers of a number field, but for simplicity only polynomial rings
are considered in this work. It is well-known that these two definitions are equivalent~\cite{RSW18}.

The RLWE problem is similar to the LWE problem, but it deals with elements in
the ring $\mathcal{R} := \Z[X]/\langle\phi(X)\rangle$ instead of a vector in
$\Z_q^n$, where $\phi(X) \in \Z[X]$ is a monic irreducible polynomial of degree $n$.
Let $s(X) \in \mathcal{R}_q := \mathcal{R} / q\mathcal{R}$
be a secret polynomial where $q \in \Z^+$.
Let $\chi_{\textup{\tiny rlwe}}$ be a probability distribution on $\mathcal{R}$
which samples $e(X) = \sum_{i=0}^{n-1} e_i X^i \in \cal R$ as $e_0,\ldots,e_{n-1} \stackrel{i.i.d.}{\longleftarrow} \chi$
using a probability distribution $\chi$ on $\Z$.
An RLWE sample is given by
\begin{align*}
\big( a(X), s(X) \!\cdot\! a(X) + e(X) \bmod q \big)
\in \mathcal{R}_q \times \mathcal{R}_q \enspace,
\end{align*}
where
$a(X) \gets \mathcal{U}(\mathcal{R}_q)$,
$e(X) \gets \chi_{\textup{\tiny rlwe}}$,
and $\cdot$, $+$ denote multiplication and addition in
$\cal R$, respectively.
The (search) RLWE problem asks to find $s(X)$,
given such RLWE samples.

\vspace{2pt}
\noindent \textit{Short integer solution.}
The SIS problem is defined as follows:
given $n$ uniformly selected random vectors
$\v a_1,\ldots, \v a_n \in \Z_q^m$ for $q \in \Z^+$,
find a nonzero $\v v = (v_1,\ldots,v_n) \in \Z^n$ such that
\begin{align*}
\sum_{i=1}^n v_i \v a_i
\equiv \v 0 \bmod q
\quad\textup{and}\quad
\norm{\v v} \le \beta \enspace,
\end{align*}
where $\sqrt{n} q^{m/n} \le \beta < q$.
The inhomogeneous version of SIS problem is to find a short
$\v v$ such that $\sum_{i=1}^n v_i \v a_i \equiv \v z \bmod q$
for given $\v z \in \Z_q^m$ where $m < n$.

\subsubsection*{\label{sec:BV}Bernstein-Vazirani algorithm}


\begin{figure*}[t]
  \centering
  \includegraphics[width=0.95\linewidth]{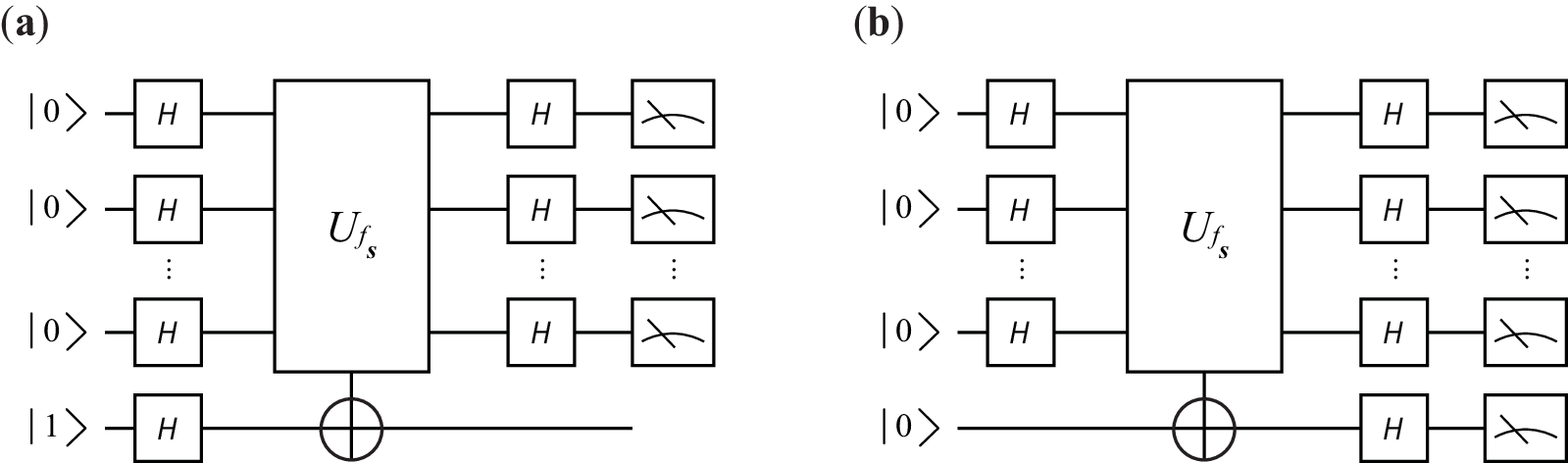}
  \caption{(\textbf{a}) A circuit description of BV algorithm and (\textbf{b}) its variant}
\label{fig:BV-circuit}
\end{figure*}

\vspace{2pt}
\noindent \textit{Binary case.}
Consider a vector $\v s \in \{0,1\}^n$ and a function
$f_\v s: \{0,1\}^n \rightarrow \{0,1\}$, $f_\v s(\v x) = \inner{\v s}{\v x} \bmod 2$.
The problem is, for given quantum oracle access to the unitary
$U_{f_\v s}:\ket{\v x} \ket{0} \mapsto \ket{\v x} \ket{f_\v s (\v x)}$,
to find $\v s$.

Figure\;\ref{fig:BV-circuit} (a) describes a quantum circuit for
the BV algorithm~\cite{DJ92,BV93}.
The initial state is $\ket{\v 0} \ket{1}$, and then applying
the gates as shown in Figure\;\ref{fig:BV-circuit} (a) achieves the goal.
Details are not covered here, but the quantum state
just before the measurement in Figure\;\ref{fig:BV-circuit} (a) reads,
\begin{align}\label{eq:BV-binary}
  \frac{1}{2^n}
  \sum_{\v y \in \{0,1\}^n}
  \sum_{\v x \in \{0,1\}^n}
  (-1)^{\inner{\v s + \v y}{\v x}}
  \ket{\v y} \ket{-} \enspace,
\end{align}
where
$\ket{-} = \frac{1}{\sqrt 2}(\ket{0} - \ket{1})$. 
As can be seen from Equation~(\ref{eq:BV-binary}), when measuring $\ket{\v y}$
register, the only non-trivial amplitude occurs for $\v y = \v s$, and
there does not exist any chance of measuring a string other than $\v s$.
Assuming single qubit gates are simultaneously executable and the query
is efficient, the time complexity of the algorithm is $O(1)$.

Slightly different circuit is introduced in Figure\;\ref{fig:BV-circuit} (b).
Using this circuit, the state before the measurement reads,
\begin{align}\label{eq:BV-binary-variant}
  \frac{1}{\sqrt 2}
  \left(
    \ket{\v 0} \ket{0}
    +
    \ket{\v s} \ket{1}
  \right) \enspace.
\end{align}
Notice that the Hadamard gate acting on the bottom wire is redundant,
but only there to make Equation~(\ref{eq:BV-binary-variant}) simple.
It can be seen from Equation~(\ref{eq:BV-binary-variant}) that
measuring the first register gives $\v s$ with probability 1/2.
The procedure can simply be repeated until we get $\v s$.

\vspace{2pt}
\noindent \textit{Generalized BV algorithm.}
Instead of qubits, consider $q$-dimensional qudits.
Each wire in Figure\;\ref{fig:BV-circuit} represents $q$-dimensional
complex vectors and Hadamard gate is accordingly generalized~\cite{qudit}.
Recall that generalized Hadamard operation can be understood as
quantum Fourier transform on $\Z_q$~\cite{bookmosca} such as
\begin{align*}
\mathsf{QFT}_{\Z_q} :
\ket{x} \mapsto
\sum_{y \in \Z_q}
e^{2 \pi \imath \frac{x y}{q}} \ket{y} \enspace .
\end{align*}

We are ready to describe the generalized BV algorithm.
Notice that after the query in Figure\;\ref{fig:BV-circuit} (b), we would have
$\frac{1}{\sqrt{q^n}} \sum_{\v x \in \Z_q^n} \ket{\v x} \ket{f_{\v s}(\v x)}$
where $f_{\v s}: \Z_q^n \rightarrow \Z_q$, $f_{\v s}(\v x) = \inner{\v s}{\v x} \bmod q$.
Instead of specifying a circuit for preparing such states,
we simply define a quantum sample $\ket{\psi} \in (\mathbb{C}^q)^{n+1}$,
and take it as an input to the BV algorithm as follows:

\begin{algorithm}[H]
  \setstretch{1}
  \caption{$\mathsf{BV}_{q,n}$}
  \label{alg:BV}
  \begin{algorithmic}[1]
  \Require $\ket{\psi}\in (\mathbb{C}^q)^{n+1}$
  \Ensure $\tilde{\v s} \in \Z_q^n$ or $\bot$
    \State Apply $\mathsf{QFT}_{\Z_q}^{n+1}$ on $\ket{\psi}$
    \State Measure all qudits to get $(\v x,y) \in \Z_q^n \!\times\! \Z_q$
    \If{$\gcd(y,q)=1$}
      \State \Return $\tilde{\v s} := -y^{-1} \v x \bmod q$
    \EndIf
    \State \Return $\bot$
  \end{algorithmic}
\end{algorithm}

\noindent One may think of the input $\ket{\psi}$ as what we get by
querying an oracle $\cal O_\v s$.
This algorithm will be the core subroutine in
the learning algorithms.

%
%

At this point, it will be a natural question to ask
if the algorithm still works for noisy input samples.
The main contribution of the previous work~\cite{GKZ19} is to answer
the question as briefly reviewed below.

\subsubsection*{LWE algorithm with quantum samples}\label{sec:qLWE}

Let $\v s \in \Z_q^n$ be a secret vector and let $\cal V \subseteq \Z_q^n$.
A quantum LWE sample for $\v s$ and $\cal V$, denoted by
$\displaystyle\ket{\psi^{\textup{\tiny lwe}}_{\v s,\cal V}}$, is defined as follows:
\begin{align}
  \ket{\psi^{\textup{\tiny lwe}}_{\v s,\cal V}} =
  \frac{1}{\sqrt{|\cal V|}}
  \sum_{\v a \in \cal V}
  \ket{\v a} \ket{\inner{\v s}{\v a} + e_\v a \bmod q}, ~
  e_\v a \gets \chi_{\textup{\tiny lwe}} \enspace ,
\label{eq:QLWE-sample}
\end{align}
where $\chi_{\textup{\tiny lwe}}$ is a probability distribution on $\Z$.
We assume that the support of $\chi_{\textup{\tiny lwe}}$ is $[-k,k] \cap \Z$  for some\
small positive integer $k$, i.e. $|e_{\v a}| \le k$.
Define a quantum oracle $\mathcal{O}^{\textup{\tiny lwe}}_{\v s,\cal V}$
which outputs $\displaystyle\ket{\psi^{\textup{\tiny lwe}}_{\v s,\cal V}}$ when queried.


Given access to the quantum oracle $\mathcal{O}^{\textup{\tiny lwe}}_{\v s,\cal V}$,
one can find the secret vector $\v s$ by using
the $\mathsf{qLWE}\textup{-}\mathsf{Solver}$ described in Algorithm~\ref{alg:qLWE} (see below).
The subroutine $\mathsf{Test}$ in Algorithm~\ref{alg:qLWE}
is designed to check if the candidate $\tilde{\v s}$
is indeed the secret vector $\v s$ by using additional LWE samples.
Note that in Algorithm~\ref{alg:Test}
the oracle $\mathcal{O}^{\textup{\tiny lwe}}_{\v s, \cal V}$ can be quantum or classical.
Detailed analysis on Algorithm~\ref{alg:qLWE} is not covered here.
Interested readers are encouraged to refer to Section IV B in the previous paper~\cite{GKZ19}.
Results and parameter choices are summarized as follows.
Probability that $\mathsf{BV}_{q,n}(\displaystyle\ket{\psi^{\textup{\tiny lwe}}_{\v s,\cal V}})$
returns the correct secret $\v s$ is lower bounded by
$\mathrm{\Omega}\!
\left( {|\cal V|}/({kq^n}) \right)$.
If $\tilde{\v s} = \v s$, $\mathsf{Test}$ returns $1$ (pass) with certainty.
If $\tilde{\v s} \ne \v s$, the probability that $1$ is returned is
at most
$  \left( ({2k \!+\! 1}) / {q} \right)^{\! \ell_2}$.
By setting $\cal V = \Z_q^n$, $\ell_1 = O(k \log (1/\eta))$, and $\ell_2 = 1$,
we have
\begin{itemize}
  \item
  success probability: $1 - \eta$
  \item
  sample complexity: $O(k \log(1/\eta))$
  \item
  time complexity: $O( k \log(1/\eta) \cdot \mathsf{poly}(n,\log q))$ \enspace.
\end{itemize}
In the last item, the time complexities of sample preparation (i.e. oracle query),
$\mathsf{BV}_{q,n}$, and $\mathsf{Test}$ are assumed to be
$O(1)$, $\mathsf{poly}(n,\log q)$, and $\mathsf{poly}(n,\log q)$,
respectively. 

\begin{algorithm}[H]
  \setstretch{1}
  \caption{$\mathsf{qLWE}\textup{-}\mathsf{Solver}$}
  \label{alg:qLWE}
  \begin{algorithmic}[1]
  \Require oracle access to $\mathcal{O}^{\textup{\tiny lwe}}_{\v s,\cal V}$
  \Ensure $\tilde{\v s} \in \Z_q^n$ or $\bot$
  %
  \State $i \gets 0$
  \Repeat
    \State $\big\vert \psi^{\textup{\tiny lwe}}_{\v s,V}\big\rangle \gets \mathcal{O}^{\textup{\tiny lwe}}_{\v s,\cal V}$
    \Comment{oracle query}
    \State $\tilde{\v s} \gets
    \mathsf{BV}_{q,n}\big(\big\vert\psi^{\textup{\tiny lwe}}_{\v s,\cal V}\big\rangle\big)$
    \If{ $\tilde{\v s} = \bot$ }
      \State \Return $\bot$
    \EndIf
    \If{ $\mathsf{Test}(\tilde{\v s}, \mathcal{O}^{\textup{\tiny lwe}}_{\v s, \cal V}) = 1$}
      \State \Return $\tilde{\v s}$
    \EndIf
    \State $i \gets i+1$
  \Until{$i \ge \ell_1$}
  \State \Return $\bot$
  \end{algorithmic}
\end{algorithm}

\begin{algorithm}[H]
  \setstretch{1}
  \caption{$\mathsf{Test}$}
  \label{alg:Test}
  \begin{algorithmic}[1]
  \Require $\tilde{\v s} \in \Z_q^n$, oracle access to $\mathcal{O}^{\textup{\tiny lwe}}_{\v s,\cal V}$
  \Ensure $1$ (pass) or $0$ (fail)
  %
  \State $i \gets 0$
  \Repeat
    \State $(\v a, b) \gets \mathcal{O}^{\textup{\tiny lwe}}_{\v s, \cal V}$
    \Comment{$b = \inner{\v a}{\v s} + e_\v a \bmod q$}
    \If{$|b - \inner{\v a}{\tilde{\v s}} \bmod q | > k$}
      \State \Return $0$
    \EndIf
    \State $i \gets i+1$
  \Until{$i \ge \ell_2$}
  \State \Return $1$
  \end{algorithmic}
\end{algorithm}

\section*{Revisiting previous works}
This section presents new results.
The first finding is about the solvability of the RLWE problem by reducing it to LWE.
The second and third findings involve the development of efficient classical algorithms
for SIS and size-reduced LWE problems under certain assumptions.

The purpose of the second and the third results is to show the following: if the assumptions on quantum samples taken in the previous works do hold, and then there also exist polynomial-time classical algorithms.
A handful of classical algorithms are introduced below, but before presenting the results, we are urged to
clarify the meaning of the word `classical' here.
In Result\;2 and Result\;3, we begin with \textit{quantum} samples assumed in respective previous works\;\cite{GKZ19,Song22a}.
We do not question or evaluate the practicality of preparing quantum samples, but rather
immediately carry out the measurement in computational basis.
Each measurement gives rise to one \textit{classical} sample that is put as an input to \textit{classical} algorithms we develop.
Therefore in estimating the complexity of classical algorithms, the number of samples used should be read as the number of quantum samples assumed.

\subsection*{\label{sec:qRLWE}Result 1: RLWE algorithm with quantum samples}
The goal of this subsection is to solve the RLWE problem given that the LWE problem is solvable by Algorithm\;\ref{alg:qLWE}.
Applying Algorithm\;\ref{alg:qLWE} to RLWE problem was originally unsuccessful by the authors of \cite[Section\;V]{GKZ19}.
Below the RLWE problem with quantum samples
is reduced to LWE by efficiently transforming a RLWE sample to an LWE sample.
The reduction utilizes linearity of operations in $\mathcal{R}$.
To be more specific, a RLWE sample is transformed to an LWE sample
in a way that the coefficients of a polynomial become a vector.
The LWE algorithm described in the previous section is then applied to recover $s(X)$.

Let $\v s(X)$ be a secret polynomial and
let $\cal V \subseteq \mathcal{R}_q$.
A quantum RLWE sample for $\v s$ and $\cal V$, denoted by
$\displaystyle\ket{\psi^{\textup{\tiny rlwe}}_{\v s,\cal V}}$ is defined as follows:
\begin{align*}
&\ket{\psi_{s,\cal V}^{\textup{\tiny rlwe}}}
= \frac{1}{\sqrt{\vert \cal V \vert}}
\sum_{a(X) \in \cal V} \ket{a(X)}\ket{s(X) \!\cdot\! a(X) + e_a(X) \bmod q}, \nn
&~~~~~~ e_a(X) \gets \chi_{\textup{\tiny rlwe}} \enspace ,
\end{align*}
where $\chi_{\textup{\tiny rlwe}}$ is a probability distribution on $\cal R$
which samples $e_a(X) = \sum_{i=0}^{n-1} e_{a,i} X^i$ as
$e_{a,0},\ldots,e_{a,n-1} \stackrel{i.i.d.}{\longleftarrow} \chi$
using a probability distribution $\chi$ on $\Z$.

\vspace{2pt}
\noindent \textit{Transform from $\cal R_q$ to $\Z_q^n$.}
Recall that the polynomial ring $\cal R$ is defined as $\cal R = \Z[X]/\langle \phi(X) \rangle$.
Let us consider a (row) vector and a matrix constructed from a polynomial $a(X) = \sum_{i=0}^{n-1} a_i X^i$ as follows:
\begin{align*}
\mathsf{V}(a) &= (a_0,a_1,\ldots,a_{n-1}) \in \Z^n, \nn
\mathsf{M}_{\phi}(a)
 &= \left(\begin{array}{ccc}
 & \mathsf{V}(a) & \\
 & \mathsf{V}(X \cdot a \bmod \phi) & \\
 & \vdots & \\
 & \mathsf{V}(X^{n-1} \cdot a \bmod \phi) &
\end{array}\right) \in \Z^{n \times n} \enspace,
\end{align*}
where $\cdot$ is a multiplication of polynomials.
Let $\phi(X) = X^n - \sum_{i=0}^{n-1} \phi_i X^i$ and define
\begin{align*}
\mathbf{P}
= \left(\begin{array}{ccccc}
0 & 1 & 0 &  \cdots & 0 \\
0 & 0 & 1 & \cdots & 0 \\
\vdots & \vdots & \vdots & \ddots & \vdots \\
0 & 0 & 0 & \cdots & 1 \\
\phi_0 & \phi_1 & \phi_2 & \cdots  & \phi_{n-1}
\end{array}\right)
\in \Z^{n \times n} \enspace .
\end{align*}
We then have a useful lemma.

\begin{lemma}\label{lem:vec-mat}
For any $a(X), b(X) \in \mathcal{R}$, the following relations hold:
\begin{align*}
\mathsf{V}(a \cdot b \bmod \phi)
= \mathsf{V}(a)\, \mathsf{M}_{\phi}(b)\,,
~~~ \mathsf{V}(X \cdot a \bmod \phi)
= \mathsf{V}(a)\, \mathbf{P} \enspace .
\end{align*}
\end{lemma}
\begin{proof}
Straightforwardly,
\begin{align*}
a(X) \cdot b(X) \bmod \phi
= \sum_{i=0}^{n-1} a_i X^i b(X) \bmod \phi
= \sum_{i=0}^{n-1} a_i \big( X^i b(X) \bmod \phi \big) \enspace,
\end{align*}
which verifies the first equation.

Next, from $\phi(X) = X^n - \sum_{i=0}^{n-1} \phi_i X^i$,
we have $X^n \equiv \sum_{i=0}^{n-1} \phi_i X^i \bmod \phi$,
and thus
\begin{align*}
X \cdot a(X)
= \sum_{i=1}^{n-1} a_{i-1} X^{i} + a_{n-1}X^n
\equiv  \sum_{i=1}^{n-1} a_{i-1} X^{i}
+ a_{n-1} \sum_{i=0}^{n-1} \phi_{i} X^{i}
\bmod \phi \enspace.
\end{align*}
Therefore,
\begin{align*}
&\phantom{=}\mathsf{V}(X \cdot a \bmod \phi)
= (0, a_0,\ldots, a_{n-2})
+ a_{n-1} (\phi_0,\phi_1,\ldots,\phi_{n-1}) \\
&= (a_0,\ldots,a_{n-1})
\left(\begin{array}{ccccc}
0 & 1 & 0 &  \cdots & 0 \\
0 & 0 & 1 & \cdots & 0 \\
\vdots & \vdots & \vdots & \ddots & \vdots \\
0 & 0 & 0 & \cdots & 1 \\
0 & 0 & 0 & \cdots  & 0
\end{array}\right)
+\  (a_0,\ldots,a_{n-1})
\left(\begin{array}{ccccc}
0 & 0 & 0 &  \cdots & 0 \\
0 & 0 & 0 & \cdots & 0 \\
\vdots & \vdots & \vdots & \ddots & \vdots \\
0 & 0 & 0 & \cdots & 0 \\
\phi_0 & \phi_1 & \phi_2 & \cdots  & \phi_{n-1}
\end{array}\right)  \\
&= \mathsf{V}(a) \, \mathbf{P} \enspace,
\end{align*}
the second equation holds.
\end{proof}

Let $\mathbf{M}^T$ denote the transpose of $\mathbf{M} \in \mathbb{R}^{l \times m}$.
Let $(j+1)$-th column vector of $\mathbf{M} \in \mathbb{R}^{l \times m}$
be denoted by $\mathbf{M}_j \in \mathbb{R}^{l \times 1}$
so that $\mathbf{M}$ can be written as
$\mathbf{M} = \left( \mathbf{M}_0 \,\vert\, \cdots \,\vert\, \mathbf{M}_{m-1} \right)$.
In particular, when $l = 1$, $\mathbf{M}_j$ denotes the $(j+1)$-th coordinate of the row vector $\mathbf{M}$.
From Lemma~\ref{lem:vec-mat},
the matrix $\mathsf{M}_\phi(b)$ can be written as
\begin{align*}
\mathsf{M}_{\phi}(b)
= \left(\begin{array}{c}
\mathsf{V}(b) \\ \mathsf{V}(b) \mathbf{P}
\\ \vdots \\ \mathsf{V}(b) \mathbf{P}^{n-1}
\end{array}\right),
\end{align*}
which shows that
\begin{align}\label{eq:column-vec}
(\mathsf{M}_\phi(b)_j)^T
= \big(
\mathsf{V}(b)_j,
(\mathsf{V}(b) \mathbf{P})_j,
\ldots, (\mathsf{V}(b) \mathbf{P}^{n-1})_j
\big)
= \big(\!
\inner{\mathsf{V}(b)}{((\mathbf{I}_n)_j \!)^T \!},
\inner{\mathsf{V}(b)}{(\mathbf{P}_{j} \!)^T \!}
\!,\ldots,\! \inner{\mathsf{V}(b)}{((\mathbf{P}^{n-1})_{j}\!)^T \!}
\!\big) \enspace.
\end{align}

\vspace{2pt}
\noindent \textit{RLWE to LWE.}
Let $\v a = \mathsf{V}(a)$ for $a\in \mathcal{R}$.
Define a unitary operation,
\begin{align*}
U_{\phi,j}: \ket{\v a}\ket{0} \mapsto \ket{\v a} \ket{\v a_j},
\quad j \in \{0,\ldots,n-1\} \enspace,
\end{align*}
where,
\begin{align*}
  \v a_j
= (\mathsf{M}_\phi(a)_j)^T
= \big(\! \inner{\v a}{((\mathbf{I}_n)_j)^T},
\inner{\v a}{(\mathbf{P}_j)^T},
\ldots,
\inner{\v a}{((\mathbf{P}^{n-1})_j)^T} \!\big) \enspace.
\end{align*}
We see that the operation is efficiently computable as any power
of $\mathbf{P}$ can be classically pre-computed by knowing the fixed polynomial $\phi(X)$.

Now let $\v s = \mathsf{V}(s)$ 
and $\v e_a = (e_{a,0},\ldots,e_{a,n-1}) = \mathsf{V}(e_a)$.
By Lemma~\ref{lem:vec-mat}, Equation~(\ref{eq:column-vec}),
and the linearity of $\mathsf{V}$, we have
\begin{align*}
\mathsf{V}(s \cdot a  + e_a \bmod \phi)
= \mathsf{V}(s) \cdot \mathsf{M}_\phi(a) + \mathsf{V}(e_a)
= \big(\! \inner{\v s}{\v a_0} + e_{a,0},
\inner{\v s}{\v a_1}  + e_{a,1},
\ldots,
\inner{\v s}{\v a_{n-1}} + e_{a,n-1} \big) \enspace.
\end{align*}
For a given quantum RLWE sample, we apply the following transformations:
\begin{align*}
\ket{a(X)} \ket{ 0}
\ket{ s(X) \!\cdot\! a(X) + e_a(X) \bmod \phi }
&\stackrel{\mathsf{V}}{\longmapsto}
\ket{\v a} \ket{\v 0} \ket{\inner{\v s}{\v a_0} + e_{a,0} ,\ldots,
\inner{\v s}{\v a_{n-1}} + e_{a,n-1} } \\
&\stackrel{U_{\phi,j}}{\longmapsto}
\ket{\v a} \ket{\v a _j}
\ket{\ldots,\, \inner{\v s}{\v a_j} + e_{a,j} ,\,\ldots}
\enspace,
\end{align*}
where $\bmod \;q$ is omitted in the third register.
Uncomputing irrelevant information,
we have a quantum sample with secret $\v s$:
\begin{align*}
\ket{\v a_j} \ket{\inner{\v s}{\v a_j} + e_{a,j} \bmod q}
\enspace.
\end{align*}

Alternatively, interchanging the role of $a(X)$ and $s(X)$ in the transformation,
we have a quantum sample with secret $\v s_j$:
\begin{align*}
\ket{a(X)} \ket{ 0}
\ket{ a(X) \!\cdot\! s(X) + e_a(X) \bmod \phi }
\stackrel{\mathsf{V}}{\longmapsto}
\ket{\v a} \ket{\v 0}
\ket{ \inner{\v a}{\v s_0} + e_{a,0} ,\ldots,
 \inner{\v a}{\v s_{n-1}} + e_{a,n-1} } \enspace,
\end{align*}
where,
\begin{align*}
  \v s_j = (\mathsf{M}_\phi(\v s)_j)^T
= \big(\! \inner{\v s}{((\mathbf{I}_n)_j)^T},
\inner{\v s}{(\mathbf{P}_j)^T},
\ldots,
\inner{\v s}{((\mathbf{P}^{n-1})_j)^T} \!\big) \enspace.
\end{align*}

In either case, the LWE algorithm is then applied to solve the problem.


\subsection*{\label{sec:qSIS}Result 2: A classical algorithm for SIS problem}
We develop an efficient classical algorithm for solving SIS given that the following quantum sample is provided.
In~\cite[Appendix B]{GKZ19},
a quantum sample for SIS is defined by,
\begin{align}\label{eq:sample-SIS}
\ket{\Psi^{\textup{\tiny sis}}}
= \frac{1}{\sqrt{q^{nm}}}
\sum_{\mathbf{A} \in \Z_q^{m \times n}}
\ket{\mathbf{A}} \ket{\mathbf{A} {\v v} \bmod q} \enspace,
\end{align}
where 
$\v v\in \Z^n$ is a short secret vector such that $0<\norm{\v v} \le \beta$.
The authors have developed a polynomial time quantum algorithm for finding
$\v v$ by using $O(2n\beta)$ quantum samples.
The detailed description of the algorithm is not covered here.
Instead, let us focus on the problem itself.
From the definition of the SIS problem,
we are given an under-determined system of equations ($m < n$).
Hardness of the SIS problem comes from the fact that among the
exponentially many solutions ($\Omega(q^{n-m})$),
there exists no known
way to choose a short one within polynomial time.

Assuming the SIS quantum sample in Equation\;(\ref{eq:sample-SIS})
is obtainable, an efficient classical algorithm can be designed.
Notice that by measuring a quantum sample, one obtains a classical sample
$(\mathbf{A},\mathbf{A} {\v v}) \in \Z_q^{m \times n} \times \Z_q^m$.
Prepare $t = \Omega(n/m)$ classical samples of the form $(\v A^{(i)},\v z^{(i)}) \in \Z_q^{m\times n} \times \Z_q^m$
which satisfies $\v A^{(i)} \v v \equiv \v z^{(i)} \bmod q$
for some nonzero vector $\v v \in \Z^n$ with $\norm{\v v} \le \beta$.
Each classical sample leads to $m$ linear equations (be reminded that one quantum sample leads to one classical sample, and one classical sample here corresponds to $m$ linear equations),
\begin{align*}
A^{(i)}_{11} v_1 + \cdots + A^{(i)}_{1n} v_n \equiv z^{(i)}_1  \bmod q,
\hspace{1mm}\ldots\hspace{1mm},\,
A^{(i)}_{m1} v_1 + \cdots + A^{(i)}_{mn} v_n \equiv z^{(i)}_m  \bmod q,
\end{align*}
where $A^{(i)}_{kl}$ is the $k$th row, $l$th column element of $\mathbf{A}^{(i)}$,
and $v_k$, $z_k^{(i)}$ are the $k$th elements of $\v v$ and $\v z^{(i)}$, respectively.
Therefore the total number of linear equations reads $mt = \Omega(n)$.
Since there are $n$ unknowns while we have $\Omega(n)$ equations,
we can find the secret vector $\v v$ by solving the system of equations using Gaussian elimination
within $O(n^3)$ time.
The total number of (quantum) samples used is $\Theta(n/m)$.

\subsection*{\label{sec:divide-conquer}Result 3: Classical algorithms for size-reduced LWE problem}
Song et al. have published a paper in 2022 which improved the result
of Grilo et al. assuming a resized sample~\cite{Song22a}.
In a nutshell, the improved algorithm and Algorithm~\ref{alg:qLWE}
work similarly except for the resized samples.
Therefore, here we drop the details on the algorithm and jump
right into the discussion on the resized sample.
Recall first that an LWE sample is
$(\v a, b = \inner{\v s}{\v a} + e \bmod q) \in \Z_q^{n} \times \Z_q$,
$\vert e \vert \le k$.

Let $\cal V_j \subseteq \Z_q$, $\v s = (s_0,\ldots, s_{n-1}) \in \Z_q^n$,
and consider the noise model $\chi_{\textup{\tiny lwe}}$
as a discrete uniform or a bounded Gaussian distribution in the interval
$[-\xi,\xi]$ around zero with $\xi \ll q$.
Assume a resized quantum sample examined by~\cite[Equation~(9)]{Song22a} is given,
\begin{align}
&    \big\vert \psi_{\cal V_j} \big\rangle
=
    \frac{1}{\sqrt{|\cal V_j|}}
    \sum_{a_j' \in \cal V_j}
    \ket{a_j'} \ket{a_j's_j + e_j' \bmod q},
\label{eq:DC-QLWE-sample}
\end{align}
where $e_j'$ is derived according to~\cite[Equation~(4)-(8)]{Song22a} satisfying,
\begin{align*}
    \vert e_j' \vert \le \xi' = O(n^3 \xi) \ll q \enspace .
\end{align*}
Below two classical algorithms are introduced assuming the resized samples are given.

\subsubsection*{Search on the error}
We first introduce an algorithm that exhaustively searches the
`correct error'.
Given a resized sample, an immediate (classical) strategy would be to
exhaustively search $s_j$ with $O(qn)$ complexity
(in which the scenario is indeed discussed in~\cite[Discussion]{Song22a}),
but a more sophisticated search would be to examine the error.

Consider an oracle $\mathcal{O}^{\textup{\tiny resized}}_{s_j, \cal V_j}$
which on query returns the (measured) resized sample of the $j$th component
$(a_j', b_j')$, where $b_j' = a_j' s_j + e_j' \bmod q$.
Given access to the oracle, Algorithm~\ref{alg:search-error} is developed.

\begin{algorithm}[H]
  \setstretch{1}
  \caption{$\mathsf{SearchError}$}
  \label{alg:search-error}
  \begin{algorithmic}[1]
  \Require error bound $\xi'$, oracle access to $\mathcal{O}^{\textup{\tiny resized}}_{s_j,\cal V_j}$
  \Ensure $\tilde{s}_j \in \Z_q$
  %
  \State $(a_j', b_j') \gets \mathcal{O}^{\textup{\tiny resized}}_{s_j, \cal V_j}$
  \State ${\rm e}\gets -\xi'$
  \Repeat
    \State $\tilde{e}_j' \gets \rm{e}$
    \State $\tilde{s}_j \gets (a_j')^{-1} (b_j' - \tilde{e}_j') \bmod q$
    \If{ $\mathsf{e\textup{-\!}Test}
    \big(\tilde{s}_j, \mathcal{O}^{\textup{\tiny resized}}_{s_j, \cal V_j}\big) = 1$}
      \State \Return $\tilde{s}_j$
    \EndIf
    \State $\rm{e} \gets \rm{e}+1$
  \Until{${\rm e} > \xi'$}
  \State \Return $1$
  \end{algorithmic}
\end{algorithm}

\begin{algorithm}[H]
  \setstretch{1}
  \caption{$\mathsf{e\textup{-\!}Test}$}
  \label{alg:eTest}
  \begin{algorithmic}[1]
  \Require $\tilde{s}_j \in \Z_q$, oracle access to $\mathcal{O}^{\textup{\tiny resized}}_{s_j, \cal V_j}$
  \Ensure $1$ (pass) or $0$ (fail)
  %
  \State $i \gets 0$
  \Repeat
    \State $(a_j'', b_j'') \gets \mathcal{O}^{\textup{\tiny resized}}_{s_j, \cal V_j}$
    \If{$|b_j'' - a_j'' \tilde{s}_j \bmod q| > \xi'$}
      \State \Return $0$
    \EndIf
    \State $i \gets i+1$
  \Until{$i \ge M$}
  \State \Return $1$
  \end{algorithmic}
\end{algorithm}

If $\tilde{e}_j' = e_j'$, then replacing $b_j'$ in Line 5 by
$a_j' s_j + e_j' \bmod q$ affirms $\tilde{s}_j = s_j$.
Since the error is exhaustively searched within the range $-\xi' \le e_j' \le \xi'$,
Algorithm~\ref{alg:search-error} eventually outputs the correct $\tilde{s}_j (= s_j)$
with time complexity $O(\xi')$, as long as the probability that wrong
$\tilde{s}_j (\ne s_j)$ passes the test is low enough.
If $\tilde{s}_j \ne s_j$, for a random sample $(a_j'', b_j'')$,
the probability that $|b_j'' - a_j'' \tilde{s}_j \bmod q| \le \xi'$
holds is $(2 \xi' + 1 )/ q$.
Therefore, such $\tilde{s}_j \ne s_j$ will pass $\mathsf{e\textrm{-}Test}$
with the probability $((2\xi'+1) / q)^M$.

In the previous work~\cite{Song22a}, the probability that the BV subroutine outputs
$\tilde{s}_j = s_j$ is $\Omega(1/\xi')$
(\cite[Equation~(17)]{Song22a} with $|v_j| \approx q$), thus $O(\xi')$ repetitions of
the subroutine would give the correct $\tilde{s}_j( = s_j)$.
The probability that their $M$-trial test accepts an incorrect answer
is at most $((2\xi' + 1)/q)^M$ as in~\cite[Equation~(12)]{Song22a}.

One may notice that the two algorithms have similar upper bounds on the
number of repetitions (of error guessing in Algorithm ~\ref{alg:search-error} and BV
subroutine in ~\cite[Section~3]{Song22a}).
Moreover, the probability that an incorrect candidate passes the
$\mathsf{e\textrm{-}Test}$ and $M$-trial test is exactly the same.
Direct comparison of the two algorithms is not immediate due to the differences
in approaches (e.g., exhaustive or else) and unit computations (quantum or classical),
but it is plausible to assess the two as similar in performance.
Except for the query, Algorithm~\ref{alg:search-error} works classically.

\subsubsection*{Lattice reduction}
There is another way to find $s_j'$ using techniques in lattice theory.
Given two classical samples
$(a_{j,k}', b_{j,k}' = a_{j,k}' s_j + e_{j,k}' \bmod q)$ for $k = 1, 2$,
consider two-dimensional vectors
$\v a_j' = (a_{j,1}', a_{j,2}')$,
$\v b_j' = (b_{j,1}', b_{j,2}')$,
$\v e_j' = (e_{j,1}', e_{j,2}')$.
Then by the construction,
we know there is a vector $\v v \in \Z^2$ such that
$\v v \equiv s_j \v a_j' \bmod q$ and
$\norm{\v b_j' - \v v} = \norm{\v e_j'}$.
In particular, the vector $\v v$ is very close to $\v b_j'$
since  $\norm{\v e_j'} \le \sqrt{2} \xi'$ is small.
Furthermore, notice that a set $\{ \v w \in \Z^2 : \v w \equiv x \v a_j' \bmod q, x \in \Z \}$
constitutes a lattice in which $\v v$ is included.
In other words, for given $\v b_j'$, finding $s_j$ can be solved
by looking for a lattice vector that is closest to $\v b_j'$,
known as the closest vector problem (CVP).

In general, the CVP is hard to solve, that is, it takes time exponential in the lattice dimension.
However in our case, it is sufficient to solve the CVP over a two-dimensional lattice,
which is tractable by using the Babai's nearest plane algorithm
with a short basis computed by the Lagrange-Gauss algorithm.

\vspace{2pt}
\noindent \textit{Lagrange-Gauss algorithm.}
This algorithm is useful for finding a shortest basis of a lattice in two-dimension.
We say that an ordered basis $\v g_1, \v g_2$ of a lattice $L$
is Lagrange-Gauss reduced if
$\norm{\v g_1} \le \norm{\v g_2} \le \norm{\v g_2 + z \v g_1}$
for all $z \in \Z$.
It can be shown that if a basis $\v g_1, \v g_2$ of a lattice $L$ is Lagrange-Gauss reduced,
then $\norm{\v g_i} = \lambda_i(L)$, where
$\lambda_i(L)$ is the $i$th successive minimum of $L$~\cite[Chapter~17]{Gal}.
This basis can be computed efficiently by Algorithm~\ref{alg:Lagrange-Gauss}.

\begin{algorithm}[H]
  \setstretch{1}
  \caption{$\mathsf{LagrangeGauss}$}
  \label{alg:Lagrange-Gauss}
  \begin{algorithmic}[1]
  \Require basis $\v g_1, \v g_2 \in \R^2$  for a lattice $L$ such that $\norm{\v g_1} < \norm{\v g_2}$
  \Ensure basis $\v g_1, \v g_2$ such that $\norm{\v g_i} = \lambda_i(L)$
  \While{$\norm{\v g_1} < \norm{\v g_2}$}
    \State $\v g_2 \gets \v g_2 - \left\lfloor\! \frac{\langle \v g_1 , \v g_2 \rangle}{\norm{\v g_1}^2} \!\right\rceil \v g_1$
    \State Swap $\v g_1$ and $\v g_2$
  \EndWhile
  \State \Return $\v g_1 , \v g_2$
  \end{algorithmic}
\end{algorithm}

\noindent
Detailed analysis of the algorithm can be found in~\cite[Chapter~17]{Gal}.
The time complexity of the algorithm is known to be $O(\log^3 (\max_i{\norm{\v g_i}^2}) )$,
where $\v g_i$ are inputs.

\vspace{2pt}
\noindent \textit{Nearest plane algorithm.}
The algorithm is for solving CVP.
Let $\v g_1, \ldots, \v g_n \in \R^n$ be basis vectors of a lattice $L$.
Given a target vector $\v t \in \R^n$, we are asked to find a vector $\v v \in L$
such that $\norm{\v t - \v v}$ is minimized.

Babai's nearest plane algorithm gives an approximate solution to CVP~\cite{Babai86}
in general dimension.
Intuitively, what it does is to fix some basis vectors and then
find a linear combination of remaining vectors such that it is closest to
the projected $\v u$ onto the hyperplane spanned by the fixed basis.
Algorithm~\ref{alg:nearest-plane-2dim} describes the nearest plane algorithm in dimension $2$.
We refer to \cite[Chapter~18]{Gal} for general dimension.


\begin{algorithm}[H]
  \setstretch{1}
  \caption{$\mathsf{NearestPlane}$}
  \label{alg:nearest-plane-2dim}
  \begin{algorithmic}[1]
  \Require basis $\v g_1, \v g_2$ for a lattice $L \subseteq \R^2$, target $\v t \in \R^2$
  \Ensure $\v v \in L$
  \State $\b g_1^\ast \gets \v g_1$
  \State $\b g_2^\ast \gets \v g_2 - \frac{\langle \v g_2, \b g_1^\ast \rangle}{\norm{\b g_1^\ast}^2} \b g_1^\ast$
  \State $\v t_2 \gets \v t$
  \State $z_2 \gets \left\lfloor\! \frac{\langle \v t_2, \b g_2^\ast \rangle}{\norm{\b g_2^\ast}^2}\!\right\rceil$
  \State $\v t_1 \gets \v t_2 - z_2 \v g_2$
  \State $z_1 \gets \left\lfloor\! \frac{\langle \v t_1, \b g_1^\ast \rangle}{\norm{\b g_1^\ast}^2} \!\right\rceil$
  \State \Return $\v v= z_1 \v g_1 + z_2 \v g_2$
  \end{algorithmic}
\end{algorithm}

\begin{proposition}\label{pro:babai}
Let $\v g_1, \v g_2 \in L$, and $\v t \in \R^2$ be inputs to Algorithm~\ref{alg:nearest-plane-2dim} and let $\b g_1^*, \b g_2^*$ be Gram-Schmidt orthogonalized vectors of $\v g_1,\v g_2$.
If there exists $\v w \in L$ such that $\norm{\v t - \v w} < \frac{1}{2} \min_{i\in \{1,2\}}(\norm{\b g_i^\ast})$,
then Algorithm ~\ref{alg:nearest-plane-2dim} outputs $\v w$.
\end{proposition}
\begin{proof}
Let $\v v = z_1 \v g_1 + z_2 \v g_2$ be the output of Algorithm~\ref{alg:nearest-plane-2dim}, and let $\v t_1, \v t_2$ be vectors as in the algorithm.
Assume there exists $\v w$ such that $\v w = w_1 \v g_1 + w_2 \v g_2 = \v t - \v e$
for some $w_1, w_2 \in \Z$ and $\norm{\v e} < \frac{1}{2} \min_{i\in \{1,2\}}(\norm{\b g_i^\ast})$.
Then we have a useful bound on $\vert\langle \v e, \b g_i^\ast \rangle\vert / \norm{\b g_i^\ast}^2$ for $i=1,2$,
\begin{align}\label{eq:err-bound}
\frac{\vert\langle \v e, \b g_i^\ast \rangle\vert}{\norm{\b g_i^\ast}^2}
\le \frac{\norm{\v e} \norm{\b g_i^\ast}}{\norm{\b g_i^\ast}^2} < \frac{1}{2} \enspace.
\end{align}
Recall that $\v t_2 = \v t = \v w + \v e$. From
\begin{align*}
&\frac{\langle \v t_2, \b g_2^\ast \rangle}{\norm{\b g_2^\ast}^2}
= \frac{\langle \v w, \b g_2^\ast \rangle}{\norm{\b g_2^\ast}^2}
+ \frac{\langle \v e, \b g_2^\ast \rangle}{\norm{\b g_2^\ast}^2}
= \frac{\langle w_1 \v g_1 + w_2 \v g_2, \b g_2^\ast \rangle}{\norm{\b g_2^\ast}^2}
+ \frac{\langle \v e, \b g_2^\ast \rangle}{\norm{\b g_2^\ast}^2}
= w_2 + \frac{\langle \v e, \b g_2^\ast \rangle}{\norm{\b g_2^\ast}^2} \enspace,
\end{align*}
we see that $z_2
= \left\lfloor \frac{\langle \v t_2, \b g_2^\ast \rangle}{\norm{\b g_2^\ast}^2} \right\rceil = w_2$
due to Equation\;(\ref{eq:err-bound}).
Similarly, recalling $\v t_1 = \v t - z_2 \v g_2$ and from
\begin{align*}
\frac{\langle \v t - z_2 \v g_2, \b g_1^\ast \rangle}{\norm{\b g_1^\ast}^2}
&= \frac{\langle \v w, \b g_1^\ast \rangle}{\norm{\b g_1^\ast}^2}
+ \frac{\langle \v e, \b g_1^\ast \rangle}{\norm{\b g_1^\ast}^2}
- z_2 \cdot \frac{\langle \v g_2, \b g_1^\ast \rangle}{\norm{\b g_1^\ast}^2} \nn
%
%
&= w_1 +
w_2 \cdot \frac{\langle \v g_2,\b g_1^\ast \rangle}{\norm{\b g_1^\ast}^2}
- z_2 \cdot \frac{\langle \v g_2, \b g_1^\ast \rangle}{\norm{\b g_1^\ast}^2}
+\frac{\langle \v e, \b g_1^\ast \rangle}{\norm{\b g_1^\ast}^2}
= w_1 +\frac{\langle \v e, \b g_1^\ast \rangle}{\norm{\b g_1^\ast}^2} \enspace,
\end{align*}
we have $z_1 = \left\lfloor \frac{\langle \v t_1, \b g_1^\ast \rangle}{\norm{\b g_1^\ast}^2} \right\rceil = w_1$,
proving $\v v = \v w$.
\end{proof}

\noindent
The time complexity of Algorithm~\ref{alg:nearest-plane-2dim} is
$O(\log^2 B )$ where
$B = \max\{\norm{\v g_1},\norm{\v g_2}\}$.

\vspace{2pt}
\noindent \textit{Finding $s_j$ by lattice reduction.}
In addition to the above algorithms, we adopt Gaussian heuristic which predicts that
the number of lattice points inside a measurable set $\cal B \subset \R^n$
is approximately equal to $\operatorname{Vol}(\cal B) / \operatorname{Vol}(L)$
where $L $ is a full-rank lattice in $\R^n$ and $\operatorname{Vol}(L) := \det(L)$.
Gaussian heuristic is widely used in analyzing the performance of lattice-related algorithms.
Applied to Euclidean ball in dimension $n$, we expect
the length of a nonzero shortest vector in a random lattice $L$ to be approximately,
\begin{align*}
  \sqrt{ \frac{n}{2\pi e}} \det (L)^{1/n} \enspace.
\end{align*}

We are now ready to describe the method.
Suppose we are given two classical samples
$(a_{j,k}' , b_{j,k}' = a_{j,k}' s_j + e_{j,k}' \bmod q)$ for $k = 1, 2$,
by measuring the resized quantum samples.
Assume that $a_{j,1}'$ is invertible in $\Z_q$.
Let $\v a = (a_{j,1}',a_{j,2}')$ and consider the following lattice:
\begin{align*}
\cal{L} =
\big\{ \v v \in \Z^2 :
\v v \equiv x \v a \bmod q
\textup{ for some } x \in \Z
\big\} \enspace,
\end{align*}
where $\bmod~ q$ applies to each component.
Notice that by letting $\v v = (v_1 , v_2) \equiv x \v a$ for some $x \in \Z$,
we have $x \equiv (a_{j,1}')^{-1} v_1 \bmod q$.
This shows that two vectors $\mathbf{b}_1 = (1, (a_{j,1}')^{-1} a_{j,2}' \bmod q)$ and
$\mathbf{b}_2 = (0, q)$ form a basis of $\cal L$, that is,
\begin{align*}
 \cal L
=
  \{ \v v:
  \v v = z_1 \mathbf{b}_1 + z_2 \mathbf{b}_2,
  ~z_1, z_2 \in \mathbb Z
  \},
\end{align*}
and thus $\dim(\cal L) = 2$, $\det(\cal L) = \det\!\big(\big[{\mathbf{b}_1 \atop \mathbf{b}_2}\big]\big) = q$.
As explained earlier, the purpose of introducing $\cal L$ is to reduce finding
$s_j$ to solving CVP:
if one manages to find $\v v \in \cal L$ that is close to
$\v b = (b_{j,1}', b_{j,2}')$, then
the error is obtained by $\v e = \v b - \v v$.

By using Lagrange-Gauss algorithm with inputs $\b b_1, \b b_2$,
we have basis vectors $\v g_1, \v g_2 \in \cal L$ such that $\norm{\v g_i} = \lambda_i(\cal L)$
for $i = 1, 2$.
Assuming the Gaussian heuristic holds,
$\lambda_1(\cal L) = \norm{\v g_1} \approx \sqrt{\det (\cal L)} = \sqrt q$.
Let $\b g_1^*, \b g_2^*$ be Gram-Schmidt orthogonalized vectors of $\v g_1,\v g_2$.
Due to the properties of Gram-Schmidt orthogonalized vectors,
we know that $\v g_1 = \b g_1^*$ and $\det(\cal L)
= \det\!\big(\big[{\mathbf{b}_1 \atop \mathbf{b}_2}\big]\big)
= \det\!\big(\big[{\b g_1^\ast \atop \b g_2^\ast}\big]\big)
= \norm{\b g_1^*} \norm{\b g_2^*}$.
Therefore, $\norm{\b g_2^*} \approx \sqrt q$.

Since $\v b \equiv s_j \v a + \v e \bmod q$,
we have $\v b = s_j \v a + \v e + q \v l$ for some $\v l \in \Z^2$
and $\norm{\v b - (s_j \v a + q\v l)} = \norm{\v e}$.
Thus, if $\norm{\v e} < \frac{1}{2}\min\{\norm{\b g_1^\ast}, \norm{\b g_2^\ast}\}
\approx \frac{1}{2}\sqrt{q}$,
then taking $\v g_1, \v g_2$ and $\v b$ as inputs,
Babai's algorithm outputs the vector $\v w = s_j \v a + q \v l$
by Proposition~\ref{pro:babai}.
Computing $\v w \bmod q$ then yields $s_j \v a \bmod q$, from which we recover $s_j$.
Indeed, since $|e_{j,i}'| \le \xi' $
(i.e., $\norm{\v e} \le \sqrt 2 \xi'$),
a condition for the algorithm to work is
\begin{align}
    8 \xi'^2 \lesssim q \enspace .
\label{eq:LG-condition}
\end{align}
In the previous paper, the authors assume $\xi' \ll q$,
thus Equation~(\ref{eq:LG-condition}) likely holds.
Considering the complexity of each subroutine algorithm,
the total running time is $\mathsf{poly}(\log(q))$.

\begin{figure*}[t]
  \centering
  \includegraphics[width=0.95\linewidth]{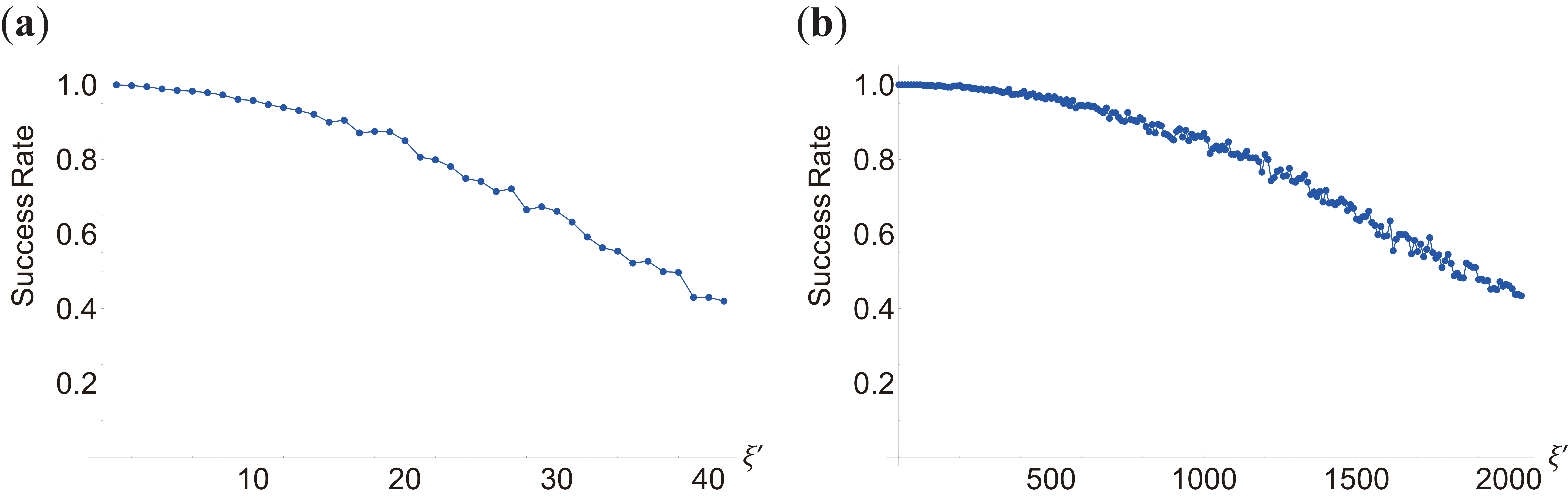}
  \caption{Success rate of the proposed algorithm as a function of the error bound for (\textbf{a}) $q=3329$ (Kyber) and for (\textbf{b}) $q=2^{23} -2^{13} + 1$ (Dilithium).
  In each figure, the condition given by Equation\;(\ref{eq:LG-condition}) is around 20 and 1023, respectively.
  In (\textbf{b}), the interval in the horizontal axis is 10.}
\label{fig:Ntest}
\end{figure*}

The above analysis relies on Gaussian heuristic which is not guaranteed to work
beneficially on every instance.
To verify the validity of the approach, numerical tests are carried out 
summarized in Figure~\ref{fig:Ntest}.
Each subfigure shows the success rate of the lattice reduction method for
chosen modulus $q$ from Kyber~\cite{kyber} and Dilithium~\cite{Dilithium}, respectively.
For each error bound $\xi'$, the following procedure is conducted 1000 times:
a pair of resized samples is produced randomly (including the secret) and the algorithm
is applied to output a candidate for the secret.
The success rate is the number of times the algorithm correctly outputs the secret
divided by 1000.
As can be seen from the figure, the algorithm succeeds better as the bound gets smaller.
There is a chance that the algorithm fails for small $\norm{\v e}$ due to unusually short
Gram-Schmidt vectors, but such cases can be overcome by getting another sample and trying
a different lattice.

\section*{Conclusion}\label{sec:conclusion}

A definitive conclusion on whether the linear learning algorithms will be immediately useful once a powerful quantum computer is ready is not yet drawn from this work, but we have shown that the RLWE problem can be tackled if LWE is indeed solvable quantumly.
In addition, it is also shown that assumptions taken in solving SIS and size-reduced LWE problems lead to efficient classical algorithms, implying that the previously claimed quantum advantages need to be reexamined.
A further multidisciplinary study on the subject may estimate the true feasibility of the algorithms.

\section*{Data availability}
All data generated or analysed during this study are included in this published article.

\bibliography{reference.bib}

\section*{Author contributions}
M.K. conceived ideas, M.K. and P.K. participated in analysis and writing the manuscript.

\section*{Additional information}

\textbf{Correspondence}
P.K. takes the role of the corresponding author.

\vspace{2pt}
\noindent\textbf{Competing interests}
The authors declare no competing interests.

\end{document}